# Reversible control of Co magnetism by voltage induced oxidation

Chong Bi[1], Yaohua Liu[2,†], T. Newhouse-Illige[1], M. Xu[1], M. Rosales[1], J.W. Freeland[3], Oleg Mryasov[4], Shufeng Zhang[1], S.G.E. te Velthuis[2], and W. G. Wang[1,*]

1) Department of Physics, University of Arizona, Tucson, Arizona 85721, USA
2) Materials Science Division, Argonne National Laboratory, Argonne, IL 60439, USA
3) Advanced Photon Source, Argonne National Laboratory, Argonne, IL 60439, USA
4) Department of Physics and Astronomy, University of Alabama, Tuscaloosa, AL 35487, USA

We demonstrate that magnetic properties of ultra-thin Co films adjacent to $Gd_2O_3$ gate oxides can be directly manipulated by voltage. The Co films can be reversibly changed from an optimally-oxidized state with a strong perpendicular magnetic anisotropy to a metallic state with an in-plane magnetic anisotropy, or to an oxidized state with nearly zero magnetization, depending on the polarity and time duration of the applied electric fields. Consequently, an unprecedentedly large change of magnetic anisotropy energy up to 0.73 erg/cm$^2$ has been realized in a nonvolatile manner using gate voltages of only a few volts. These results open a new route to achieve ultra-low energy magnetization manipulation in spintronic devices.

* wgwang@physics.arizona.edu
† yhliu@anl.gov

It has been a long sought-after goal to control the magnetic properties of solids by electric fields (EF) [1–6], with the premise that voltage-induced magnetization reversal can be much more energetically efficient than that of magnetic fields or spin-polarized currents [7]. Much of the effort has been focused on multiferroic materials, where intrinsic correlation exists among the magnetic, electric, and elastic orders [1,2,5]. Substantial progress has also been made in magnetic semiconductors, where the magnetism can be controlled by EF-dependent charge carrier density [3,4]. However, multiferroic materials and magnetic semiconductors usually lack the desired properties of common 3d ferromagnets and their alloys such as large magnetization and high spin polarization, and most of them only function at low temperatures, which limits their applications in spintronic devices [8,9].

Recently, it has been discovered that EFs applied through a liquid electrolyte could modify the switching fields of perpendicularly magnetized FePt and FePd thin films [10]. This work has triggered an intense interest in studying EF-controlled magnetic properties in 3d FMs and their alloys [11–20]. It turns out that the finite penetration length of EF in 3d FMs, of the order of ~1 Å [21], can have a large impact when the magnetic anisotropy has an interfacial origin. This effect of voltage-controlled magnetic anisotropy (VCMA) has been successfully employed to achieve low switching current density in magnetic tunnel junctions [12], to induce sub-ps precessional switching [13,14], to change the ferromagnetic ordering temperature [15] and to substantially modify domain wall propagation velocity [16]. The perpendicular magnetic anisotropy energy in these systems is related to the nondegenerate in-plane and out-of-plane d orbitals due to the hybridization of 3d orbitals of FM and the 2p orbitals of oxygen at the FM/oxide interface [22]. Theoretically, these VCMA effects have been understood by the change of electron density among different d orbitals of FMs in the presence of EFs [17–20]. Although very rapid progress has been made in this subject, the largest VCMA effect achieved today is only about 30-50 fJ/Vm [12–14,23]. Moreover, like many other magnetoelectric effects, previous VCMA lacks nonvolatility because the change of magnetic anisotropy vanishes when the EF is turned off. Very recently it has been demonstrated that a strong nonvolatile domain wall trap can be realized by EFs on the perimeter of patterned Pt/Co/GdO$_x$ structures due the migration of $O^{2-}$ driven by voltage [24], which could lead to large anisotropy change [25] and induce reversible oxidation [26]. This type of EF-induced ion displacement has also been studied in memristors [27,28] and correlated insulators [29] where the motion of $O^{2-}$ can lead to large change of resistance up to a few orders of magnitude.

Here we report, for the first time, the giant voltage controlled magnetism (VCM) of Co thin films adjacent to $Gd_2O_3$ gate oxides. Unlike the previously reported VCMA effects, we show that both the saturation magnetization ($M_S$) and anisotropy field ($H_A$) of the Co layers can be simultaneously controlled by EFs in a nonvolatile fashion, resulting in a record high change of magnetic anisotropy energy up to 0.73 erg/cm$^2$ with gate voltages of only a few volts, equivalent to an electric field effect of 11.6 pJ/Vm. Through a combination of structural, magnetic, transport and spectroscopic studies, we have demonstrated that this giant VCM effect is achieved by voltage-induced reversible oxidation of the Co layer, which can be understood by a large interfacial EF and the high $O^{2-}$ ion mobility in $Gd_2O_3$.

The nominal structure of the samples in this study is Si/SiO$_2$/Pt(4nm)/Co(0.7nm)/Gd$_2$O$_3$(80nm)/Ta(5nm)/Ru(100nm). For transport measurements, the samples were patterned into Hall bar structures with a feature width of 2.5 μm. In order to determine the oxidation states and the magnetization of Co layers, we have performed X-ray magnetic circular dichroism (XMCD) measurements using beamline 4-ID-C at the Advanced Photon Source. Details of sample fabrication and measurement are provided in the supplementary Material [30].

Fig. 1(a) shows the schematic of the sample structure and the transport measurement geometry. We used the anomalous Hall effect (AHE) [31] to characterize the magnetic properties of patterned samples after the application of EFs. In order to better illustrate the effect of voltage-controlled magnetism in this system, EFs were applied at elevated temperatures ranging between 200°C and 260°C, then all the transport measurements were conducted with zero EF after samples were cooled to room temperature (RT).

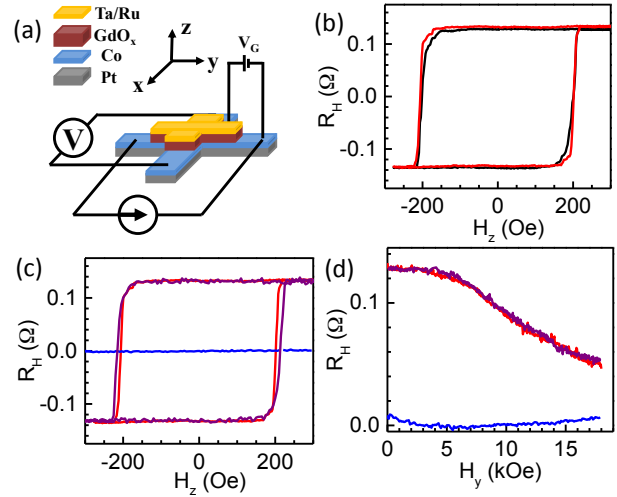

Fig. 1. (Color online) (a) Schematic of the sample structure and experimental setup. (b) $R_H$-$H_z$ curves of the sample in as-deposited state (black) and after staying at 200 °C for 10 min (red). (c) $R_H$-$H_z$ curves of the sample after the application of EF = -625 kV/cm for 6 min (blue) and EF = +625 kV/cm for 13 min (purple) at 200°C. The red curve is the same as in (b). (d) The corresponding $R_H$-$H_y$ curves for the three cases shown in (c).

We first measured the anomalous Hall resistance $R_H$ as a function of a perpendicular external field ($H_z$) for the as-deposited sample. As shown in Fig. 1(b), the square $R_H$-$H_z$ curve indicates that the virgin sample shows a strong perpendicular magnetic anisotropy (PMA) with the coercive field ($H_c$) of 200 Oe and the amplitude of $R_H$ of 0.27 Ω. The AHE curve of the sample after staying at 200 °C for 10 mins without any EF applied is also plotted in Fig. 1(b). The two curves closely resemble each other, demonstrating heating alone has little impact on the sample's magnetic properties. Surprisingly, the AHE hysteresis loop nearly disappears after a small EF of -625 kV/cm (corresponding to a gate voltage of -5V) is applied as shown in Fig. 1(c). The $R_H$-$H_z$ curve now only shows a very weak ordinary Hall signal, which doesn't saturate at high $H_z$ (not shown), and the $R_H$-$H_y$ curve measured with an in-plane magnetic field also vanishes as shown in Fig. 1(d), suggesting the vanish of $R_H$ is not due to in-plane anisotropy. These facts indicate that the applied negative EF has a profound influence on the magnetism of the Co layer. Since the AHE curves were measured after the EF was turned off, the observed nonvolatile behavior here is obviously not due to the charge transfer effect [12–14]. Instead, it implies that the change is caused by EF-driven ion motion. Rear-earth oxides such as $Gd_2O_3$ are known as an ionic conductor with very high $O^{2-}$ mobility [24,32]. As one may expect, a negative EF will drive $O^{2-}$ towards Co layer therefore dramatically altering the magnetic properties of Co.

Very interestingly, the original PMA can be completely restored. After the negative EF, a positive EF of 625 kV/cm was applied for 13 mins at 200 °C. Subsequently, $R_H$ curves were measured at RT. As shown in Fig. 1(c), the $R_H$-$H_z$ curve is almost fully recovered to its initial shape. The longitudinal resistance of the Hall bars only changed less than 5% after the entire process. The same $R_H$-$H_z$ curves indicate both the $M_z$ and $H_c$ were recovered after the positive EF. Moreover, the $R_H$-$H_y$ curve under the in-plane field is also completely restored as shown Fig. 1(d). This hard-axis AHE curve directly links to the perpendicular anisotropy field of Co layer. Its recovery after the positive EF further confirms the PMA has been reversibly changed to its initial state.

In the pioneering studies on PMA of Pt /Co/AlOx trilayers, it was found that strong PMA was associated with an optimally oxidized Co/AlOx interface, whereas in-plane magnetic anisotropy was observed in under-oxidized samples and PMA with multidomain structure was observed in over-oxidized samples [33,34]. Here we demonstrated that the same effect can be achieved *in-situ* by the applied gate voltage in a single sample in a reversible manner. More importantly, in contrast to the over-oxidized samples in previous studies, the $R_H$-$H_y$ curves of our samples nearly vanish after negative EFs, suggesting the continuous Co film has been turned into very small superparamagnetic Co islands embedded in $CoO_X$, or even has been totally oxidized by the migrated $O^{2-}$ driven by applied voltages.

At this point, we would like to summarize below three striking features of the observed VCM effect. First, the degree of magnetic property change is truly giant when compared to the charge-transfer-induced VCMA effects [12–14,23]. Vibrating sample magnetometry (VSM) studies on these perpendicularly magnetized Pt/Co/$Gd_2O_3$ films showed a saturation magnetization of 1200 emu/cm$^3$ and an anisotropy field of 12.5 kOe (see Supplementary Material [30]); this translates to an effective surface perpendicular energy density, $K^\perp \cdot t$ ($K^\perp$ = ½ $M_S \cdot H_A$ and $t$ is the thickness of the film), of 0.53 erg/cm$^2$, controlled entirely by a small EF of 625 kV/cm. By comparison, a large EF of ~10 MV/cm is required for a change of ~0.02 erg/cm$^2$ in the Ta/CoFeB/MgO system [12–14,23]. Second, this giant magnetism control is reversible; this may seems counter-intuitive since the chemical reaction near FM/oxide interface is typically to be an irreversible process [33–36]. Our results, including additional experiments shown below, demonstrate that the ionic migration and subsequent chemical reaction processes in Co/ $Gd_2O_3$ are essentially reversible. Finally, the VCM effect is nonvolatile; this is in a sharp contrast with the conventional VCMA in which the effect comes from EF-induced electron density redistribution and thus intrinsically suffers volatility [12–14,23].

It is known that a substantial amount of interfacial $CoO_x$ is crucial to the strong PMA in Pt/Co/AlOx [33,34].To further confirm our discovery of giant EF-controlled magnetism, a direct link between the amount of interfacial $CoO_x$ modulated by the EF and the strength of the PMA is desirable. Next, the EFs were applied at a moderately higher temperature of 260 °C. The evolution of magnetic properties of the Pt/Co/$Gd_2O_3$ trilayer is shown in Fig. 2. We start with the zero magnetization state created by a negative EF as shown in Fig. 2(a). Upon application of +625 kV/cm for only 30 s, the AHE curve nearly returns to the as-deposited state. After the application of the positive EF for 120 s, the AHE curve exhibits a larger $H_C$ and larger $R_H$ compared to the as-deposited state, indicating the Co film now has a stronger PMA. Note the absence of exchange-bias behavior is expected here due to the low Neel temperature of the very thin CoO layer. $Hc$ of the Co layer keeps increasing with further application of a positive EF until 150 s, after which $H_C$ starts to decrease, accompanied with the decrease of $R_H^R$ / $R_H^S$ ( $R_H^R$ and $R_H^S$ are the remanent and saturated Hall resistance, respectively). Very interestingly, $R_H^S$ keeps increasing in the entire duration of positive EF application. After 600 s, the AHE curve exhibits a hard-axis hysteresis loop under the out-of-plane field, as shown in Fig. 2(d), with a $R_H^S$ being nearly doubled that of the as-deposited state and a negligible Hc. Finally, Figs. 2(e) and (f) show that the Co film can be restored to the PMA state and subsequently to the initial state by the negative EF in a much shorter time scale. The observation of a hard-axis like AHE curve in Fig. 2(d) is significant. $R_H$ in this state doesn't reach saturation until $H_z$ = 3000 Oe. This fact, together with the nearly zero $R_H^R / R_H^S$ ratio, indicates that the easy axis of Co film has been turned to the in-plane orientation, which is supported by the previous study showing that a Co film in under-oxidized Pt/Co/AlOx structures had an in-plane easy axis [33,34].

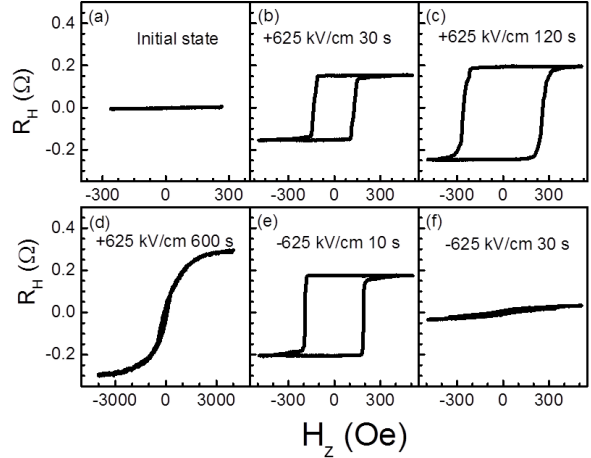

Fig. 2. EF-induced evolution of magnetism of a Co layer measured by AHE. (a) The AHE curve for the initial state with $R_H \sim$ 0 Ω. (b-d) AHE curves after EF = + 625 kV/cm for different durations, and (e-f) AHE curves after EF = -625 kV/cm for different durations. EF was applied at 260 °C and all AHE curves were recorded at RT.

In order to measure the $M_s$ as well as the oxidation state of Co films, we have carried out XMCD experiments in the fluorescence yield (FY) mode before and after the application of EFs. An applied magnetic field of 3.5 kOe was applied perpendicular to the film plane during the measurements. Note that to measure $M_s$ with large samples of a few mm$^2$, that are normally required by SQUID or VSM, is unfeasible due to large probability of defect-assisted dielectric breakdown. The Co $L_3$ edge absorption peak of the sample in the as-deposited state is shown in Fig. 3(a). In addition to the main Co peak at 778.6 eV, a shoulder is clearly visible at 779.8eV, indicative of an interfacial CoO$_x$ layer between Co and Gd$_2$O$_3$ that is expected for samples with strong PMA [33,34]. The normalized XMCD signal at the $L_3$ edge is shown in Fig. 3(b). The total magnetic moment per Co atom calculated from the sum rule is 0.92±0.10 $\mu_B$ (see Supplementary Material [30]), which is reasonably close to 1.05±0.10 $\mu_B$ determined from the VSM measurement. The sample shows a completely different behavior after the application EF = -625 kV/cm for 10 min at 260 °C. The shoulder at 779.8 eV has turned into a peak. Two other peaks at 777.3 eV and 782.2 eV, characteristic of CoO [37], start to emerge, making the spectrum almost identical to that of CoO as previously reported [38]. Stronger evidence of a Co$^{2+}$-dominating state is that the peaks at 778.6 eV and 779.8eV are nearly of the same height, consistent with a loss of metallic Co in the film [38,39]. At the same time, there is no detectable magnetic signal in the XMCD spectrum as shown in Fig. 3(b). These facts indicate that no isolated Co particles remain, and almost the entire Co film has been oxidized into CoO by the negative EF driven O$^{2-}$ migration. Remarkably, the sample shows nearly metallic behavior after the application of positive EF under the same condition. Now the Co $L_3$ peak is much narrower with no shoulders and the peak position is consistent with metallic Co. The total moment per Co atom determined from the XMCD measurement is 1.65±0.10 $\mu_B$, very close to the value of pure Co (1.6 $\mu_B$) [40], demonstrating the in-plane easy axis observed in Fig 2(d) indeed was due to the formation of metallic Co by the application of positive EF.

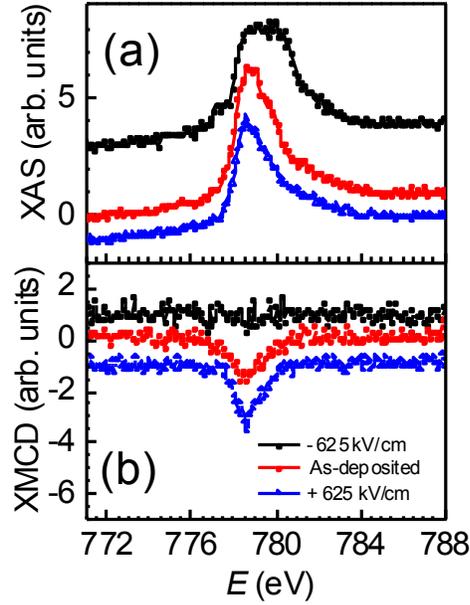

Fig. 3. Normalized XAS spectra (a) and XMCD spectra (b) at the Co $L_3$ edge showing the EF-controlled oxidation state and magnetization of ultra-thin Co films. The curves have been vertically shifted for clarity.

With these results, it is estimated that Co/Gd$_2$O$_3$ system can be changed between an in-plane state with -$K^{//}\cdot t$ =0.20 erg/cm$^2$ , where $K^{//}$ is calculated from the anisotropy field determined from Fig. 2(d) and the magnetization measured by XMCD, to a strong PMA state with $K^{\perp}\cdot t$ =0.53 erg/cm$^2$, reaching a total surface anisotropy energy density change of 0.73 erg/cm$^2$ only by a small EF of 625 kV/cm, equivalent to a VCMA effect of 11.6 pJ/Vm. This giant control of magnetism is much larger than the charge-trap induced anisotropy change [41,42] and more than two orders of magnitude larger than the normal VCMA effect of 30-50 fJ/Vm in CoFeB/MgO or Fe/MgO [12–14,23].

The dynamic behavior of the VCM effect at different temperatures is shown in Fig. 4. For positive EF, the initial state of the sample is consisted of CoO as shown in Fig. 2(a), while for negative EF the initial state is with in-plane anisotropy as shown in Fig. 2(d). Thermally activated behavior is evident from the very different time scales at different temperatures. While it takes 3000 s to achieve $R_H^S$ = 0.3 Ω with +625 kV/cm at 200 °C, it only takes 30s at 260 °C. $H_c$ reaches a maximum around 105s with +625 kV/cm at 260 °C, and starts to decline after this point due to the decrease of $R_H^R / R_H^S$. Note this maximum value of $H_c$ could not be achieved at lower temperatures under the amount of time explored. Clearly the control of magnetism under a negative EF is much faster than that for positive EF. For example it takes 600 s to change CoO state to Co state with +625 kV/cm at 260 °C; but it only takes 30 s to return to the nearly fully oxidized state. This difference is likely related with the asymmetric CoO/Gd$_2$O$_3$ barrier and the additional energy required for nucleation of metallic Co.

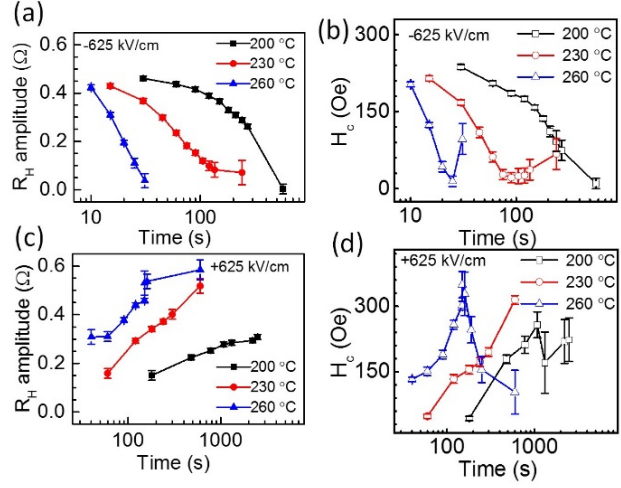

Fig. 4. Time dependence of the amplitude of $R_H$ and $H_c$ for applying (a, b) negative EF and (c, d) positive EF at different temperatures.

The observed VCM may be qualitatively explained by the following model. One can readily estimate the voltage dropped within the $Gd_2O_3$ layer, $V_{Gd_2O_3}$ and the voltage across the $Co/CoO_x$ interface layer, $V_{int}$, by utilizing the boundary condition at the interface [43],

$$\varepsilon_0 \frac{V_{int}}{\lambda} = \varepsilon \frac{V_{Gd_2O_3}}{L}. \quad (1)$$

where $\lambda$ is the thickness of $CoO_x$ layer including the screening length of metallic Co layer and $L$ is the thickness of $Gd_2O_3$ layer. By using $V_{in} + V_{Gd_2O_3} = V_g$ along with the above relation, we find the voltage drop across the interface is

$$V_{int} = \frac{\varepsilon_r \lambda V_g}{\varepsilon_r \lambda + L} \quad (2)$$

If we take $\varepsilon_r = 22$ [44], $\lambda = 0.4$ nm, $L$=80nm, and $V_g$ = 5 V, we have $V_{int}$= 0.5 V, which generates an interfacial EF that is nearly 20 times larger than the average EF of 625 kV/cm. With this significant voltage across the interface, the mobile oxygen ion near the interface is able to overcome the electronic barrier of $CoO_x$ so that the thermally assisted oxygen ion transport becomes possible. Reversibly, when the bias voltage is positive, the oxygen ion in $Co/CoO_x$ would move into $Gd_2O_3$ in the presence of the positive electric field. While this interpretation provides plausible explanation to the observed VCM effect, several issues should be addressed in order to fully understand the physics involved. Among them, the dielectric constant, interface oxygen concentration, and EF dependence of $Co/CoO_x$ interface, should be carefully characterized before a quantitative picture can be used to explain the dynamic behaviors shown in Fig.4.

In summary, we have demonstrated a VCM effect where the magnetism of ultra-thin Co films can be controlled by voltage-driven reversible oxidation with a strong asymmetric behavior for EFs with different polarities. A giant change of magnetic anisotropy energy up to 0.73 erg/cm$^2$ has been achieved by applying a small voltage of a few volts, which can be qualitatively understood by a large interfacial EF and the high $O^{2-}$ mobility of the gate oxide. These results open a new pathway to achieve voltage-controlled spintronic devices by directly manipulating the magnetism, rather than only the magnetic anisotropy, of 3$d$ transitional FMs.

**Acknowledgement**
This work was supported in part by NSF (ECCS-1310338) and by C-SPIN, one of six centers of STARnet, a Semiconductor Research Corporation program, sponsored by MARCO and DARPA. Work at Argonne National

Laboratory (Y.L. and S.G.E.t.V.) was supported by the US Department of Energy, Office of Science, Basic Energy Sciences, Materials Sciences and Engineering Division. Work performed at the Advanced Photon Source was supported by the US Department of Energy, Office of Science, Office of Basic Energy Sciences, under contract no. DE-AC02-06CH11357.

# Supplemental Information

**Table of Contents:**

**1. Sample fabrication**
**2. Structural characterization**
**3. Magnetic properties of continuous film**
**4. XMCD measurements**
**5. Comparison study with CoFeB/MgO**
**6. Technological importance of the voltage-controlled magnetism effect**

**1. Sample fabrication**

The Co/$Gd_2O_3$ films were deposited on thermally oxidized silicon substrate by dc magnetron sputtering at a base pressure of $5\times10^{-8}$ torr. The deposition rate is less than 0.5 Å s$^{-1}$ for all layers. The continuous film structure is Si/$SiO_2$/Pt(4)/Co(0.7)/$Gd_2O_3$(80)/Ta(5)/Ru(100), where the number inside parentheses is thickness in unit of nm. $Gd_2O_3$ layer was deposited by reactive sputtering from a metal Gd target. A metal Gd layer ~1 nm was first deposited before introducing oxygen into sputtering chamber, which is a common method to avoid oxidation of under layers. By using standard photolithography and ion beam etching, samples were patterned into Hall bar structures with a feature width of 2.5 μm. Subsequently Ta(5)/Ru(100) gate electrodes were patterned as shown in Fig. 1(a). In the micro-fabrication process, a 100 nm $SiO_2$ layer was deposited immediately after each etching process before breaking the vacuum, so all edges of the samples were protected by $SiO_2$ layer and were not exposed in atmosphere.

In order to confirm the observed large effect is related to $Gd_2O_3$, we have investigated the behavior of perpendicular MgO MTJ under the same conditions. The blanket films were fabricated by sputtering with a base pressure of $6\times10^{-9}$ torr with structure of Si/$SiO_2$/Ta(5nm)/Ru(10nm)/Ta(5nm)/ $Co_{20}Fe_{60}B_{20}$(0.9nm)/ MgO(1nm)/ $Co_{20}Fe_{60}B_{20}$(1.5nm)/Ta(10nm)/Ru(24nm). The MgO layer was deposited by RF sputtering under a pressure of 1.1

mTorr, with deposition rate ~ 0.2 Å/s The metal layers were deposited by DC sputtering under a pressure of 2 mTorr. The films were subsequently patterned into circular MTJs with diameters ranging from 2μm-20μm.

## 2. Structural Characterization

We have performed X-ray diffraction experiment (The Philips X'Pert MPD) to determine the crystal structure of samples. The film structure is Si/SiO$_2$/Pt(4nm)/Co(0.7nm)/GdO$_x$(80nm). The diffraction pattern is shown in Figure S1. The Gadolinium oxide films deposited at room temperature show a crystalline structure instead of being amorphous [1]. The two peaks at 2θ = 20.4° and 2θ = 28.8° can be indexed as the (211) and (222) diffraction peaks of cubic Gd$_2$O$_3$ with a lattice constant of 10.8 Å. The Pt buffer layer is <111> oriented evident from the strong (111) peak at 39.5° and the (222) peak at 85.2°. The broad peak at ~69.2° is due to the Si substrate, which is not observed for films deposited on a glass substrate.

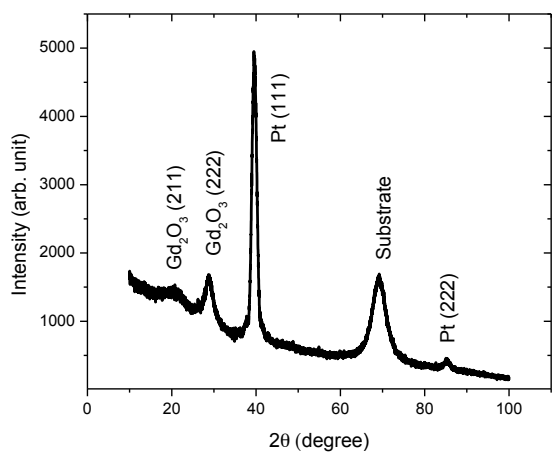

Figure S1. X-ray diffraction pattern for Si/SiO$_2$/Pt(4nm)/Co(0.7nm)/GdO$_x$(80nm).

## 3. Magnetic properties of continuous films

Figure S2. shows hysteresis loops of a continuous film measured by vibrating sample magnetometer (Microsense, EZ9 series) with magnetic fields applied in the in-plane and perpendicular-to-plane orientation. This plot shows that the saturation magnetization and perpendicular anisotropy field are about 1200 emu/cm$^3$ and 12.5 kOe, respectively. The background signals from the substrate and sample holder have been subtracted.

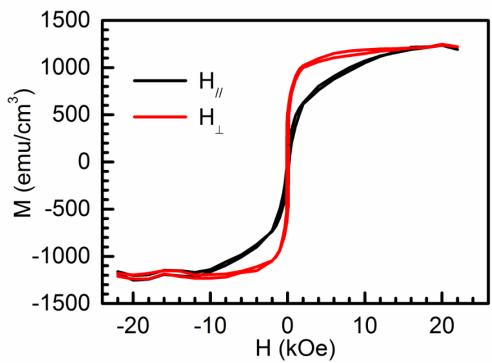

Figure S2. Hysteresis loops of continuous film (as-deposited) measured by VSM.

## 4. X-ray Absorption and X-ray Magnetic Circular Dichroism.

Figure S3 shows the X-ray Absorption (XAS) and X-ray Magnetic Circular Dichroism (XMCD) spectra of three patterned samples with the size of 200 x 200 µm². The XMCD experiments were conducted at beam line 4-ID-C at the Advanced Photon Source [2]. Circularly polarized x rays were used to obtain absorption spectra recorded by fluorescence yield. The X-ray spot was focused to be 150 x 150 µm². The XMCD (XAS) spectra are given by the difference (sum) between the absorption spectra of the right and left circularly polarized x-rays. Data were collected by tuning the X-ray energy at Co $L_{2,3}$ edge, with an incident beam angle of 70° with respect to the film plane and with an applied magnetic field of 3.5 kOe perpendicular to the film plane. The data are normalized by the x-ray intensity and then the edge jump before and after the $L_3$ edge of the average absorption spectra.

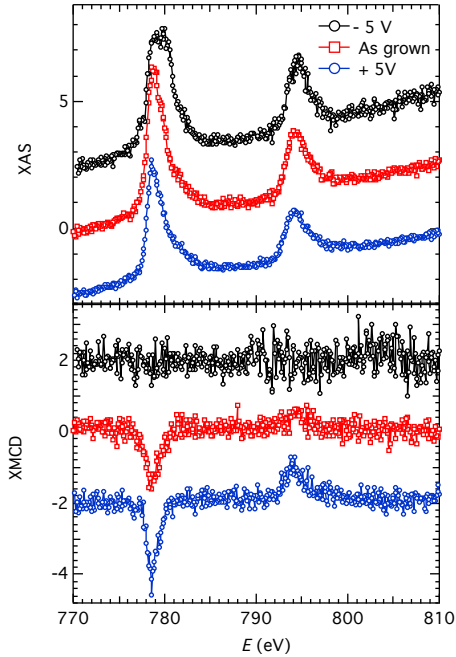

Figure S3. XAS and XMCD spectra for Co films before and after the application of EF with different polarities, which provide the element-specific information on valence holes and magnetism.

The intensity sum rule states that the transition intensity is proportional to the valence hole number $n_h$ when we sum over the $L_{2,3}$ peak [3]. Using this rule, we found that the ratio of 3d hole per Co atom is 1:1.24:1.38 for the +5 V, the as-grown and the -5 V samples, respectively.

The spin and orbit moments of the Co magnetization were quantitatively estimated using the XMCD sum-rule in units of $\mu_B$/atom [4]:

$$m_{orb} = -n_h \frac{4 \int_{L3+L2}(\mu_+ - \mu_-)dE}{3 \int_{L3+L2}(\mu_+ + \mu_-)dE}$$

$$m_{spin} = -n_h \left(1 + \frac{7 <T_z>}{2 <S_z>}\right)^{-1} \frac{2 \int_{L3}(\mu_+ - \mu_-)dE - 4 \int_{L2}(\mu_+ - \mu_-)dE}{3 \int_{L3+L2}(\mu_+ + \mu_-)dE}$$

, where µ+( µ-) is the absorption intensity with left (right) circular polarized x-rays, $n_h$ is the number of holes in the d shells. The factor $\frac{7<T_z>}{2<S_z>}$ is typically very small (a fraction of 1 %) for metallic Co films, and thus can be ignored [4]. The photon incident angle (70°) has been taken into account by multiplying by [1/cos(70°)]. The XAS spectrum of the +5V sample is almost identical to metallic Co [4]. The reported values of *3d* hole number in metallic Co are between 2.5 and 2.8 per atom for metallic Co [3]. Using these values, the lower and upper limits of the spin (orbital moment) are calculated to be 1.61 (-0.05) $\mu_B$ and 1.80 (-0.05) $\mu_B$ per Co atom, respectively, for the + 5 V sample. For the fresh sample, the lower and upper limits of the spin (orbital moment) are calculated to be 0.74 (0.13) $\mu_B$ and 0.83 (0.14) $\mu_B$ per Co atom, respectively. The -5 V sample did not show appreciable magnetization.

The direct comparison between the reference spectra of CoO and Co [5] and that of our sample after the application of bias voltage are shown in Figure S4. For both the CoO obtained after the application of negative voltage and pure Co obtained after the application of positive voltage in our sample, the spectra closely resemble that of the references.

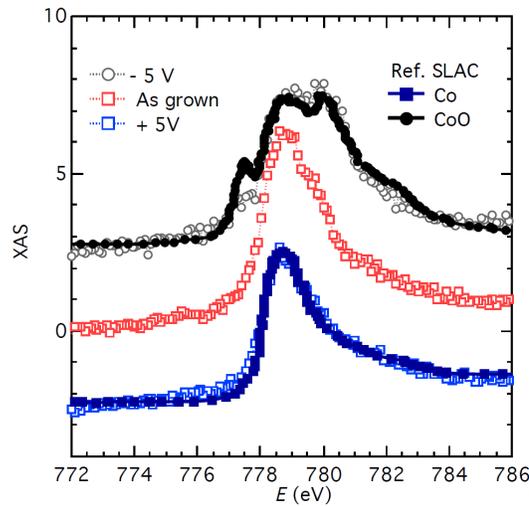

Figure S4. Comparison between the reference CoO and Co spectra with the Pt/Co/Gd2O3 structure after the application of bias voltages

## 5. Comparison study with CoFeB/MgO

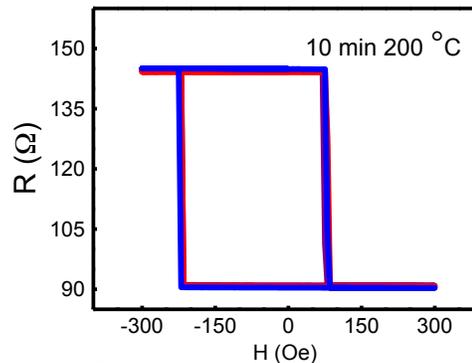

Figure S5. TMR curves of a CoFeB/MgO tunnel junction in the virgin state (blue), after EF = -625 kV/cm for 10 min at 200 °C(red), and after EF = +625 kV/cm for 10 min at 200 °C(purple).

Oxygen ion electromigration was proposed theoretically as one of possible sources of VCMA in Fe/MgO [6]. For comparison, we have measured a CoFeB/MgO/CoFeB MTJ after the same process as shown in Figure S5. EFs of ±625 kV/cm were applied to the MTJ for 10 min at 200 °C. Then the tunneling magnetoresistance curve ( minor loop) was measured after the MTJ was cooled down to room temperature. The three TMR curves totally overlap with each other, demonstrating that the large change of magnetism is due to high ion mobility in $Gd_2O_3$ and this effect is very small [7] in MgO-based MTJs within the electric field and temperature ranges we have explored.

### 6. Technological importance of the VCM effect

Finally we would like to comment on the technological importance of the VCM effect. Although the time scale (seconds) in the present study is large, it is possible to dramatically reduce this by optimizing the FM/oxide interfacial structures, considering the resistance change due to the voltage-driven $O^{2-}$ motion can be very fast (ns) as demonstrated in memristors [8,9]. The high temperature required here ( 200 °C - 260 °C) is also well within the accessible range in heat assisted magnetic recording [10] or thermally-assisted MRAM switching [11]. At the same time, the applied bias voltage (5V) can be further decreased with thinner $Gd_2O_3$ or by using other oxides with higher $O^{2-}$ mobility. Another important point is that the lattice constant of the cubic $Gd_2O_3$ obtained here has only a 5% mismatch to CoFe. Therefore it is possible to realize this VCM effect in CoFe(001)/ $Gd_2O_3$ (001)/CoFe(001) MTJs with large magnetoresistance.